\documentstyle[11pt,moriond,epsf]{article}
\newlength{\listwidth}
\def\be{\begin{equation}}
\def\ee{\end{equation}}
\def\bea{\begin{eqnarray}}
\def\eea{\end{eqnarray}}

\newcommand{\nub}{\overline{\nu}}
\newcommand{\cbar}{\overline{c}}

\begin{document}
\vspace*{1cm}
\title{
MEASUREMENTS OF $F_2$, $xF_3^{\nu}$-$xF_3^{\nub}$
FROM CCFR $\nu_\mu$-Fe and $\nub_\mu$-Fe
DATA IN A MODEL INDEPENDENT WAY 
}

\author{ 
A.~Bodek,$^{7}$ U.~K.~Yang,$^{7}$ T.~Adams, $^{4}$ A.~Alton,$^{4}$
C.~G.~Arroyo,$^{2}$ S.~Avvakumov,$^{7}$ L.~de~Barbaro,$^{5}$
P.~de~Barbaro, $^{7}$ A.~O.~Bazarko,$^{2}$ R.~H.~Bernstein,$^{3}$
 T.~Bolton, $^{4}$ J.~Brau,$^{6}$ D.~Buchholz,$^{5}$
H.~Budd,$^{7}$ L.~Bugel,$^{3}$ J.~Conrad,$^{2}$ R.~B.~Drucker,$^{6}$
B.~T.~Fleming,$^{2}$
J.~A.~Formaggio,$^{2}$ R.~Frey,$^{6}$ J.~Goldman,$^{4}$
M.~Goncharov,$^{4}$ D.~A.~Harris,$ ^{7} $ R.~A.~Johnson,$^{1}$
J.~H.~Kim,$^{2}$ B.~J.~King,$^{2}$ T.~Kinnel,$ ^{8}$
S.~Koutsoliotas,$^{2}$ M.~J.~Lamm,$^{3}$ W.~Marsh,$^{3}$
D.~Mason,$^{6}$ K.~S.~McFarland, $^{7}$ C.~McNulty,$^{2}$
S.~R.~Mishra,$^{2}$ D.~Naples,$^{4}$ P.~Nienaber,$^{3}$
A.~Romosan,$^{2}$ W.~K.~Sakumoto,$^{7}$ H. Schellman,$^{5}$
F.~J.~Sciulli,$^{2}$ W.~G.~Seligman,$^{2}$ M.~H.~Shaevitz,$^{2}$
W.~H.~Smith,$^{8}$ P.~Spentzouris,$^{2}$ E.~G.~Stern,$^{2}$
M.~Vakili,$^{1}$ A.~Vaitaitis,$^{2}$ V.~Wu,$^{1}$
J.~Yu,$^{3}$ G.~P.~Zeller,$^{5}$ and E.~D.~Zimmerman$^{2}$\\
(The CCFR/NuTeV  Collaboration)}

\address{
$^{1}$ Univ. of Cincinnati, Cincinnati, OH 45221;
$^{2}$ Columbia University, New York, NY 10027 \\
$^{3}$ Fermilab, Batavia, IL 60510
$^{4}$ Kansas State University, Manhattan, KS 66506 \\
$^{5}$ Northwestern University, Evanston, IL 60208;
$^{6}$ Univ. of Oregon, Eugene, OR 97403 \\
$^{7}$ Univ. of Rochester, Rochester, NY 14627;
$^{8}$ Univ. of Wisconsin, Madison, WI 53706\\
Presented by Arie Bodek at QCD Moriond, March 2000: hep-ex/0005021; UR1605}



\maketitle\abstracts{
We report on the extraction of  the structure functions
$F_2$ and  $\Delta xF_3 = xF_3^{\nu}$-$xF_3^{\nub}$
from CCFR  $\nu_\mu$-Fe and $\nub_\mu$-Fe differential cross sections.
The extraction is performed in a physics model independent (PMI) way.
This first measurement for  $\Delta xF_3$, which is useful in testing
models of heavy charm production, is higher
than current theoretical predictions.
The $F_2$ (PMI) values  measured in $\nu_\mu$ and $\mu$ scattering
are in good agreement with the predictions of
Next to Leading Order PDFs (using massive charm production 
schemes), thus resolving the long-standing discrepancy
between the two sets of data.}
%
%
%
\section{Introduction}
 Deep inelastic lepton-nucleon scattering experiments have been used
to determine the quark distributions in the nucleon.
However, the quark distributions determined from muon
and neutrino
experiments were found to be different at small values of $x$,
because of a disagreement in the extracted structure functions.
Here,
we report on a measurement of differential cross sections
and structure functions from CCFR $\nu_\mu$-Fe and $\nub_\mu$-Fe data.
We find that the neutrino-muon difference is resolved by
extracting the $\nu_\mu$ structure functions in a physics
model independent way.

The sum of $\nu_\mu$ and $\nub_\mu$
 differential cross sections 
for charged current interactions on an isoscalar target is related to the
structure functions as follows:
\begin{tabbing}
$F(\epsilon)$ \= $\equiv \left[\frac{d^2\sigma^{\nu }}{dxdy}+
\frac{d^2\sigma^{\overline \nu}}{dxdy} \right] 
 \frac {(1-\epsilon)\pi}{y^2G_F^2ME_\nu}$
 $ = 2xF_1 [ 1+\epsilon R ] + \frac {y(1-y/2)}{1+(1-y)^2} \Delta xF_3 $.
\end{tabbing}
Here $G_{F}$ is the Fermi weak coupling constant, $M$ is the nucleon
mass, $E_{\nu}$ is the incident energy, the scaling
variable $y=E_h/E_\nu$ is the fractional energy transferred to
the hadronic vertex, $E_h$ is the final state hadronic
energy, and $\epsilon\simeq2(1-y)/(1+(1-y)^2)$ is the polarization of 
the virtual $W$ boson.
The structure
function $2xF_1$ is expressed in terms of $F_2$
by $2xF_1(x,Q^2)=F_2(x,Q^2)\times
\frac{1+4M^2x^2/Q^2}{1+R(x,Q^2)}$, where $Q^2$ is the
square of the four-momentum transfer to the nucleon,
  $x=Q^2/2ME_h$ (the Bjorken scaling variable) is
the fractional momentum carried by the struck quark,
and $R=\frac{\sigma
_{L}}{\sigma _{T}} $ is the ratio of the cross-sections of
longitudinally- to transversely-polarized $W$-bosons.
The $\Delta xF_3$ term, which in leading order
 $\simeq 4x(s-c)$, is not present in the $\mu$-scattering case.
In addition, in a  $\nu_\mu$ charged current interaction with
$s$ (or $\cbar$) quarks, there is a threshold suppression originating
from the production of heavy $c$ quarks in the final state.
For $\mu$-scattering, there is no suppression for scattering
from $s$ quarks, but more suppression when scattering
from $c$ quarks since there are two heavy
 quarks ($c$ and $\cbar$) in the final state.  

In previous analyses
of $\nu_\mu$ data,
light-flavor universal
physics model dependent (PMD) structure functions were extracted
by applying a slow rescaling correction to correct for the charm
mass suppression in the final state. In addition, 
the $\Delta xF_3$ term (used as input in the extraction)
was calculated from a leading order charm production model. These
resulted in a physics model dependent (PMD) structure functions. 
 In the
new analysis reported here, slow rescaling corrections are not applied,
 and $\Delta xF_3$ and $F_2$
are extracted from two parameter fits to the data.
We compare the values of $\Delta xF_3$
to various charm production models.
The extracted physics model independent (PMI) values for  $F_2^{\nu}$
are then compared  with $F_2^{\mu}$
within the framework of NLO models for massive charm production.

\section{Results}

The CCFR experiment  collected  data
using the Fermilab Tevatron Quad-Triplet wide-band  $\nu_\mu$ and $\nub_\mu$
beam.
The raw differential cross sections per nucleon on iron
 are determined in bins of $x$, $y$, 
and $E_{\nu}$ ($0.01 < x < 0.65$, $0.05<y<0.95$, and $30< E_\nu <360$. GeV). 
Figure \ref{fig:dxf3} (a)
 shows typical differential cross sections 
at $E_\nu=150$ GeV.
Next, the raw
 cross sections are corrected for electroweak radiative
effects,
 the $W$ boson propagator, and for the
 5.67\% non-isoscalar excess 
of neutrons over protons in iron (only important at high $x$).
Values of $\Delta xF_3$ and $F_2$ are extracted from the sums of
the corrected $\nu_\mu$-Fe and $\nub_\mu$-Fe 
differential cross sections at different
energy bins according to Eq. (1).
It is challenging to fit $\Delta xF_3$, $R$, and $2xF_1$ 
using the $y$ distribution at a given $x$ and $Q^2$
because of the  strong correlation 
between the $\Delta xF_3$ and $R$ terms, unless the full range of
$y$ is covered by the data.
Covering this range 
(especially the high $y$ region) is hard
because of the low acceptance. Therefore, we restrict
the analysis to two parameter fits.

Our strategy is  to fit $\Delta xF_3$ and $2xF_1$ (or equivalently $F_2$)
for $x<0.1$ where the $\Delta xF_3$ contribution is relatively large, while
constraining $R$ using the
$R_{world}^{\mu/e}$
QCD inspired empirical fit
to all available $R$ from electron- and $\mu$-scattering data.
The $R_{world}^{\mu/e}$ fit is also in good agreement
with NMC $R^\mu$ data
at low $x$, and
with the most recent NNLO QCD calculations (including target mass effects)
of $R$ by Bodek and Yang

For $x<0.1$,  $R$ in neutrino scattering is expected to be somewhat larger
than $R$ for muon scattering because 
of the production of massive charm quarks
in the final state.  A correction for this
difference is applied to $R_{world}^{\mu/e}$ using a
 leading order slow rescaling model
to obtain an effective $R$ for neutrino scattering, $R_{eff}^{\nu}$. The
difference between  $R_{world}^{\mu/e}$ and $R_{eff}^{\nu}$
 is used as a systematic
error.  Because of the positive correlation between
$R$ and $\Delta xF_3$, the extracted values of $F_2$ are rather insensitive
to the input $R$. If a large input $R$ is used, a larger value of $xF_3$
is extracted from the $y$ distribution, thus yielding the
same value of $F_2$. In contrast, the extracted values of $\Delta xF_3$
are sensitive to the assumed value of $R$, which is reflected in
a larger systematic error. 
The values of $\Delta xF_3$ are sensitive to the energy 
dependence of the neutrino flux ($\sim$ $y$ dependence),
 but are insensitive to the absolute normalization.
The uncertainty on the flux shape is estimated by using the 
constraint that $F_2$
and $xF_3$ should be flat over $y$ (or $E_{\nu}$) for each $x$ and $Q^2$ bin.

\begin{minipage}{0.5\textwidth}
\epsfxsize=3.0in \epsfbox{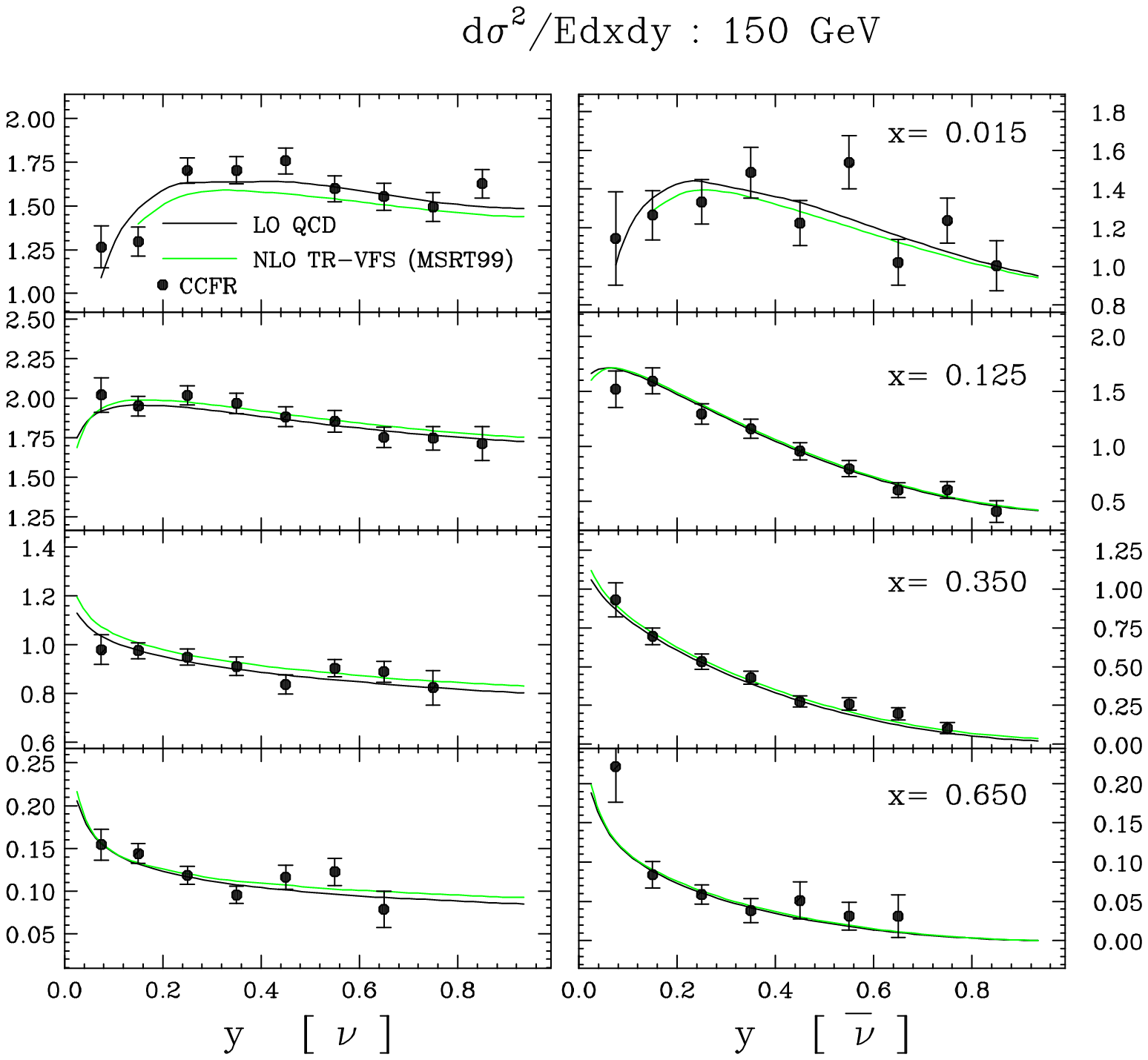}
\end{minipage}
\begin{minipage}{\listwidth}
\epsfxsize=3.0in \epsfbox{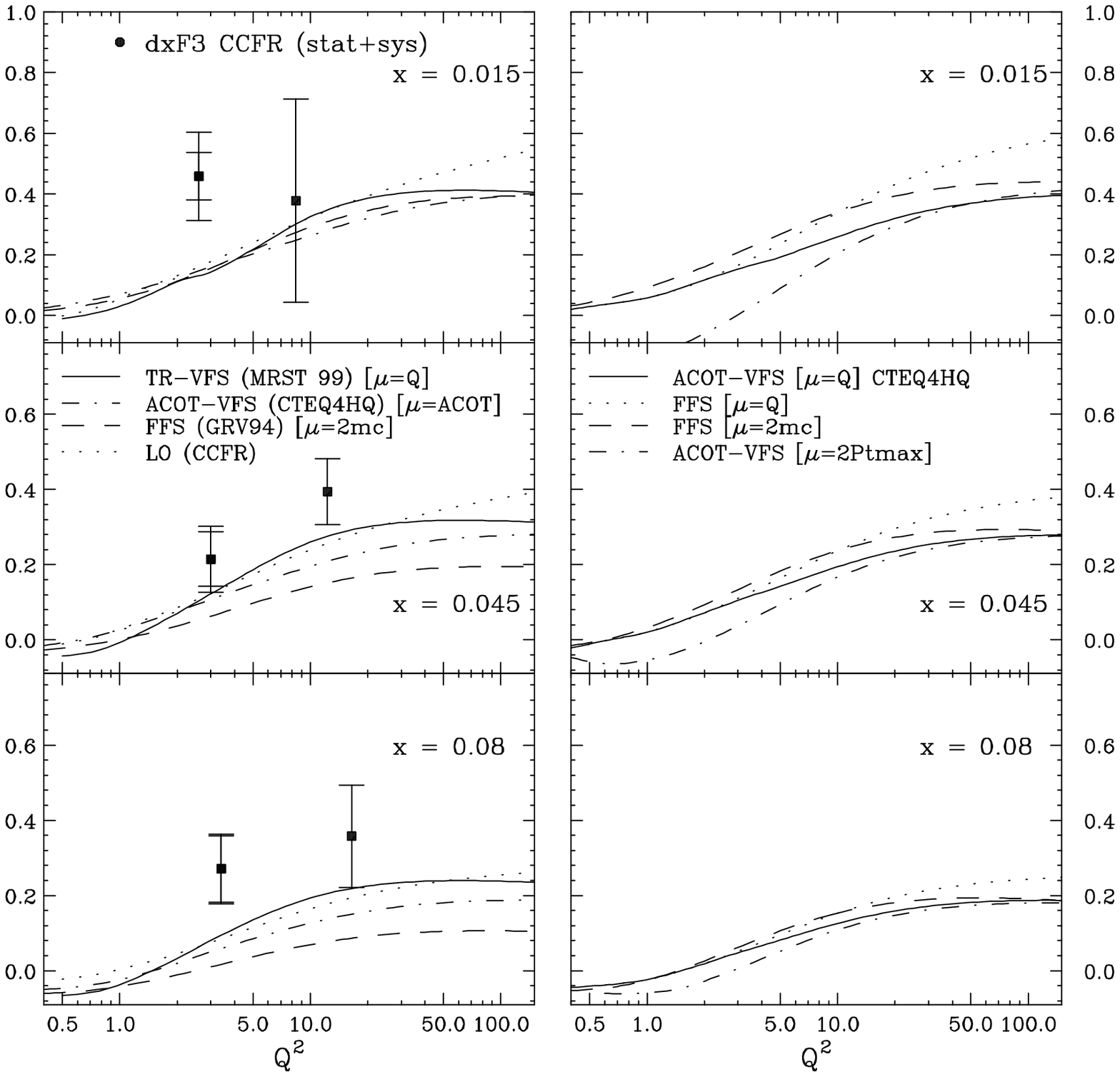}
\end{minipage}
\vspace{-0.2in}
\begin{figure}[h]
\caption{
(a) Typical raw differential cross sections at $E_\nu=150$ GeV
(both statistical and systematic errors are included).
(b) $\Delta xF_3$ data as a function of $x$
compared with various schemes for massive charm production:
 RT-VFS(MRST),
 ACOT-VFS(CTEQ4HQ), FFS(GRV94), and LO(CCFR), a
leading order model with a slow rescaling correction (left);
Also shown is the sensitivity of the theoretical calculations
to the choice of scale (right).}
\label{fig:dxf3}
\end{figure}

Because of the limited statistics, we use 
large bins in $Q^2$ in the extraction of $\Delta xF_3$ with bin centering
corrections from  the NLO
Thorne \& Roberts Variable Flavor Scheme (TR-VFS)
calculation with the MRST PDFs.
Figure~\ref{fig:dxf3} (b)  shows the extracted
values of $\Delta xF_3$ as a function of $x$, including both statistical
and systematic errors, compared to
various theoretical methods for modeling
heavy charm productions within a QCD framework. 
The three-flavor Fixed Flavor Scheme (FFS)
 assumes that there is no intrinsic charm 
in the nucleon, and all scattering from
$c$ quarks occurs via the gluon-fusion diagram.  The concept behind the
Variable Flavor Scheme (VFS) proposed by ACOT
is that at low
scale, $\mu$, one uses the three-flavor FFS scheme, and above some scale,
one changes to a four-flavor calculation and an intrinsic
charm sea (which is evolved from zero) is introduced. The concept in
the RT-VFS scheme
is that it starts with the three-flavor FFS scheme
at a low  scale,
becomes the four-flavor VFS scheme at high scale, and
interpolates smoothly between the two regions.
Shown are the predictions from the
TR-VFS scheme (as corrected after DIS-2000 and implemented
with MRST PDFs),
 with their suggested scale
 $\mu=Q$, and the predictions of the
other two NLO calculations,
ACOT-VFS (implemented with CTEQ4HQ
and the recent ACOT
 suggested scale $\mu = m_c$ for $Q<m_c$, and
$\mu^2=m_c{^2}+cQ^2(1-m_c{^2}/Q^2)^n$ for $Q<m_c$ with $c=0.5$ and $n=2$), and
the FFS (implemented with the GRV94
PDFs and
GRV94 recommended scale
$\mu = 2m_c$).
Also shown are the predictions from
 $\Delta xF_3 \simeq 4Ks(x,Q^2)$
from a leading order model (LO(CCFR))
Buras-Gaemers type fit to the CCFR dimuon
data (here K is a slow rescaling correction).
Figure~\ref{fig:dxf3} (b) (right) also shows the sensitivity to
the choice of scale.
The data do not favor the ACOT-VFS(CTEQ4HQ) predictions
if implemented with an earlier suggested scale of
$\mu=2Pt_{max}$.
With reasonable choices of scale, all the theoretical models
yield similar results. However,
at low $Q^2$  our $\Delta xF_3$ data are higher than all
the theortical models.
The difference between data and theory
may be due to an underestimate
of the strange sea (or gluon distribution)
 at low $Q^2$, or from missing
NNLO terms.

As discussed above, values of $F_2$ (PMI) for $x<0.1$ are extracted from
two parameter fits to the $y$ distributions.
In the $x>0.1$ region,
the contribution from  $\Delta xF_3$ is small and the extracted
values of $F_2$ are insensitive to $\Delta xF_3$. Therefore,
we extract values of $F_2$ with an input value of $R$
and with $\Delta xF_3$ constrained to the TR-VFS(MRST) predictions. 
As in the case of the two parameter fits for $x<0.1$, no 
corrections for slow rescaling are applied. 
Fig.~\ref{fig:f2} (a) shows our $F_2$ (PMI)
measurements divided by the predictions
from the TR-VFS(MRST) theory. Also shown are $F_2^{\mu}$ and
$F_2^e$ from the
NMC
  divided
by the theory predictions. In the calculation of the QCD
 TR-VFS(MRST) predictions, we have also included corrections for nuclear
effects,
 target mass and
higher twist corrections
 at low values of $Q^2$. As seen in
Fig.~\ref{fig:f2}, both the CCFR and NMC
structure functions are in good agreement with the TR-VFS(MRST) predictions,
and therefore in good agreement with each other.  A comparison
using the ACOT-VFS(CTEQ4HQ) predictions yields similar results.

In the previous analysis
 of the CCFR data, the extracted values of $F_2$ (PMD) at the lowest
 $x=0.015$ and $Q^2$ bin
were up to $20\%$ higher than both the NMC data and
the predictions of the light-flavor
MRSR2 PDFs.
(see figure ~\ref{fig:f2} (b) ). 
 About half
 of the difference originates from having used a leading
order model for $\Delta xF_3$ versus using our new measurement. 
The other half originates
from having used the leading order slow rescaling corrections, instead of
using a NLO massive charm production model,
and from improved modeling of the low $Q^2$ PDFs (which
changes the radiative corrections and the overall absolute normalization
to the total neutrino cross sections).

\begin{minipage}{0.5\textwidth}
\epsfxsize=3.0in \epsfbox{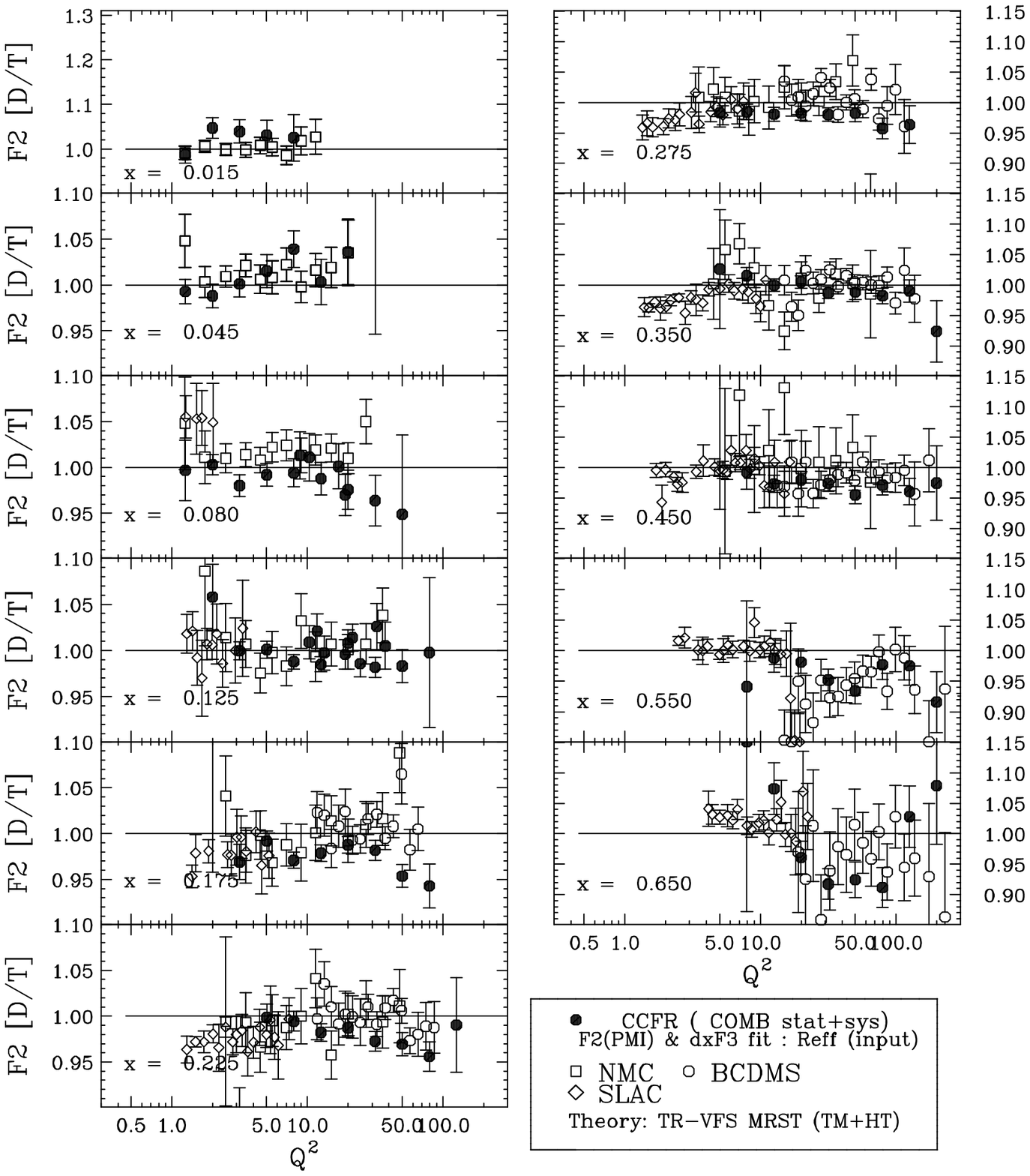}
\end{minipage}
\begin{minipage}{\listwidth}
\epsfxsize=3.0in \epsfbox{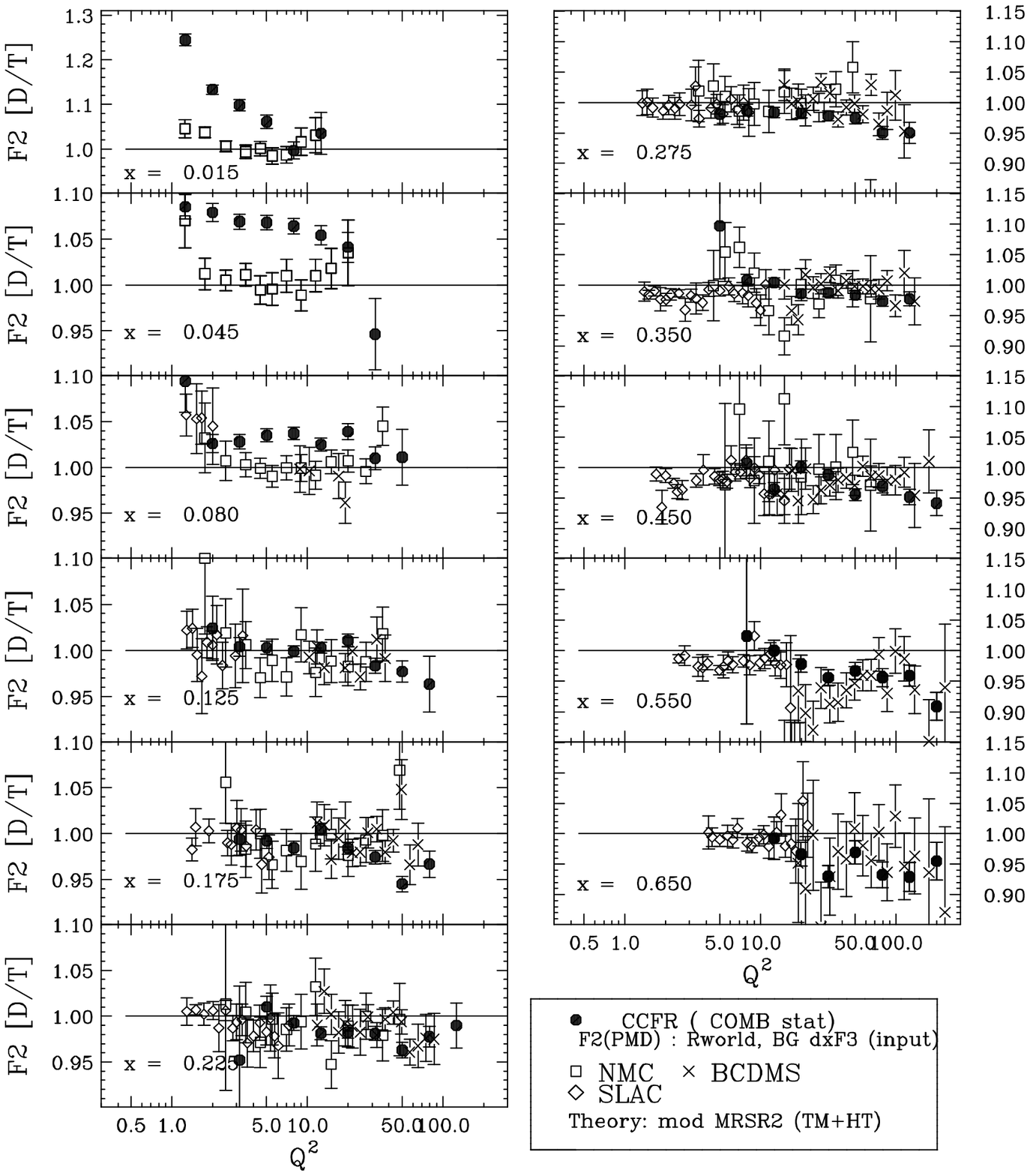}
\end{minipage}
\vspace{-0.2in}
\begin{figure}[h]
\caption{
(a) Left side: The ratio (data/theory) of the $F_2^{\nu}$ (PMI)
data divided by the predictions of TR-VFS(MRST) (with
target mass and higher twist corrections).
Both statistical and systematic errors are included. Also shown
are the ratios of
the $F_2^{\mu}$ (NMC) and $F_2^e$ (SLAC) to the TR-VFS(MRST)
predictions. (b) Right side: The ratio (data/theory) of the previous
$F_2^{\nu}$ (PMD) data (and also $F_2^{\mu}$ (NMC) and $F_2^e$ (SLAC))
divided by the predictions of the MRSR2 light-flavor PDFs (with
target mass and higher twist corrections).
}
\label{fig:f2}
\end{figure}

\section{Conclusions}

In conclusion, the $F_2$ (PMI) values
measured in neutrino-iron and muon-deuterium scattering
show good agreement with with the predictions of
Next to Leading Order PDFs (using massive charm production 
schemes), thus
resolving the long-standing discrepancy between the two
sets of data. The first measurements of $\Delta xF_3$ are higher
than current theoretical predictions.

\end{document}